%% file: main.tex
\pdfoutput=1
\documentclass[12pt,a4paper]{article}

\usepackage{ifthen} 
\newboolean{pdflatex}
\setboolean{pdflatex}{true} 

\newboolean{articletitles}
\setboolean{articletitles}{true} 

\newboolean{uprightparticles}
\setboolean{uprightparticles}{false}

\def\paperauthors{LHCb collaboration} 
\def\paperasciititle{Test of lepton flavour universality with Bs -> phi ell ell decays} 
\def\papertitle{Test of lepton flavour universality with $B_s^0 \rightarrow \phi \ell^+\ell^-$ decays} 
\def\paperkeywords{{High Energy Physics}, {LHCb}} 
\def\papercopyright{\the\year\ CERN for the benefit of the LHCb collaboration} 
\def\paperlicence{CC BY 4.0 licence}
\def\paperlicenceurl{https://creativecommons.org/licenses/by/4.0/}

\newif\ifEnableSectionTOCLinks
\EnableSectionTOCLinksfalse

\input{preamble}
\usepackage{longtable}
\usepackage{booktabs}
\usepackage{overpic}

\def\BsPhiMM   {\decay{\Bs}{\phi\mup\mun}} 
\def\BsPhiEE   {\decay{\Bs}{\phi\ep\en}} 
\def\BsPhiLL   {\decay{\Bs}{\phi\ellell}}
\def\BsJpsiPhi {\decay{\Bs}{\jpsi\phi}} 
\def\BsJpsiPhiMM    {\decay{\Bs}{\jpsi(\to \mumu) \phi}} 
\def\BsJpsiPhiEE    {\decay{\Bs}{\jpsi(\to \ep\en) \phi}} 
 
\def\BsPsiTwoSPhi   {\decay{\Bs}{\psitwos\phi}}

\def\rphi      {\ensuremath{R_\phi}\xspace}
\def\invrphi   {\ensuremath{R_\phi^{-1}}\xspace}
\def\rjpsi     {\ensuremath{r_\jpsi}\xspace}

\def\Rpsitwos  {\ensuremath{R_\psitwos}\xspace}
\def\mKKLL     {\ensuremath{m(\Kp\Km\ellell)}\xspace}
\def\mKKMM     {\ensuremath{m(\Kp\Km\mumu)}\xspace}
\def\mKKEE     {\ensuremath{m(\Kp\Km\ep\en)}\xspace}
\def\zfit      {\textsc{zfit}\xspace}

\def\invgevgevcccc {\ensuremath{\gev^{-2}c^{4}}\xspace}

\begin{document}

\renewcommand{\thefootnote}{\fnsymbol{footnote}}
\setcounter{footnote}{1}

\input{title-LHCb-PAPER}
\cleardoublepage

\renewcommand{\thefootnote}{\arabic{footnote}}
\setcounter{footnote}{0}

\pagestyle{plain} 
\setcounter{page}{1}
\pagenumbering{arabic}

\input{body}

\input{acknowledgements}

\clearpage
\input{appendix}

\clearpage
\addcontentsline{toc}{section}{References}
\bibliographystyle{LHCb}
\bibliography{main,standard,LHCb-PAPER,LHCb-CONF,LHCb-DP,LHCb-TDR}
  
\newpage
\input{Authorship_LHCb-PAPER-2024-032}

\end{document}

%% file: preamble.tex
\usepackage[top=1in, bottom=1.25in, left=1in, right=1in]{geometry}

\usepackage{microtype}
\usepackage{lineno}  
\usepackage{xspace} 
\usepackage{caption}

\usepackage{graphicx}
\usepackage{color}
\usepackage{colortbl}
\graphicspath{{./figs/}} 

\usepackage{amsmath} 
\usepackage{amssymb}
\usepackage{amsfonts}
\usepackage{upgreek}

\newcommand*\patchAmsMathEnvironmentForLineno[1]{%
\expandafter\let\csname old#1\expandafter\endcsname\csname #1\endcsname
\expandafter\let\csname oldend#1\expandafter\endcsname\csname
end#1\endcsname
 \renewenvironment{#1}%
   {\linenomath\csname old#1\endcsname}%
   {\csname oldend#1\endcsname\endlinenomath}%
}
\newcommand*\patchBothAmsMathEnvironmentsForLineno[1]{%
  \patchAmsMathEnvironmentForLineno{#1}%
  \patchAmsMathEnvironmentForLineno{#1*}%
}
\AtBeginDocument{%
\patchBothAmsMathEnvironmentsForLineno{equation}%
\patchBothAmsMathEnvironmentsForLineno{align}%
\patchBothAmsMathEnvironmentsForLineno{flalign}%
\patchBothAmsMathEnvironmentsForLineno{alignat}%
\patchBothAmsMathEnvironmentsForLineno{gather}%
\patchBothAmsMathEnvironmentsForLineno{multline}%
\patchBothAmsMathEnvironmentsForLineno{eqnarray}%
}

\usepackage{hyperxmp}

\usepackage[pdftex,
            pdfauthor={\paperauthors},
            pdftitle={\paperasciititle},
            pdfkeywords={\paperkeywords},
            pdfcopyright={Copyright (C) \papercopyright},
            pdflicenseurl={\paperlicenceurl}]{hyperref}
\usepackage[bottom,flushmargin,hang,multiple]{footmisc}

\usepackage[all]{hypcap} 

\input{lhcb-symbols-def}

\hypersetup{
  colorlinks   = true, 
  urlcolor     = blue, 
  linkcolor    = blue, 
  citecolor    = red   
}

\ifEnableSectionTOCLinks
    \usepackage[explicit]{titlesec}
    
    \let\oldcontentsline\contentsline
    \renewcommand

    \titleformat{\section}{\normalfont\Large\bf}{\hyperlink{tocsection.\thesection}{{\thesection} \parbox[t]{\dimexpr\textwidth-1pc}{#1}}}{1pc}{}

    \titleformat{\subsection}{\normalfont\bf}{\hyperlink{tocsubsection.\thesubsection}{{\thesubsection} \parbox[t]{\dimexpr\textwidth-1pc}{#1}}}{1pc}{}

    \titleformat{name=\section,numberless}[display]{}{}{0pt}{\normalfont\Huge\bfseries #1}
\fi

\usepackage{cite}
\usepackage{mciteplus}

%% file: lhcb-symbols-def.tex
\usepackage{xspace} 
\usepackage{upgreek}

\def\lhcb   {\mbox{LHCb}\xspace}

\def\cms    {\mbox{CMS}\xspace}

\def\babar  {\mbox{BaBar}\xspace}
\def\belle  {\mbox{Belle}\xspace}

\def\MagUp {\mbox{\em Mag\kern -0.05em Up}\xspace}

\ifthenelse{\boolean{uprightparticles}}%
{

 \def\Pmu         {\ensuremath{\upmu}\xspace}                 
 \def\Pnu         {\ensuremath{\upnu}\xspace}                 
                  
 \def\Ppi         {\ensuremath{\uppi}\xspace}

 \def\Ppsi        {\ensuremath{\uppsi}\xspace}

 \def\PDelta      {\ensuremath{\Delta}\xspace}                 
 \def\PXi         {\ensuremath{\Xi}\xspace}                 
 \def\PLambda     {\ensuremath{\Lambda}\xspace}                 
 \def\PSigma      {\ensuremath{\Sigma}\xspace}                 
 \def\POmega      {\ensuremath{\Omega}\xspace}                 
 \def\PUpsilon    {\ensuremath{\Upsilon}\xspace}
 \let\oldPi\Pi
 \def\PPi         {\ensuremath{\oldPi}\xspace}

 \def\PB      {\ensuremath{\mathrm{B}}\xspace}                 
 \def\PD      {\ensuremath{\mathrm{D}}\xspace}                 
 \def\PJ      {\ensuremath{\mathrm{J}}\xspace}                 
 \def\PK      {\ensuremath{\mathrm{K}}\xspace}                 
 \def\Pb      {\ensuremath{\mathrm{b}}\xspace}                 
 \def\Pc      {\ensuremath{\mathrm{c}}\xspace}                 
                  
 \def\Pe      {\ensuremath{\mathrm{e}}\xspace}                 
 \def\Pp      {\ensuremath{\mathrm{p}}\xspace}                 

 \def\Ps      {\ensuremath{\mathrm{s}}\xspace}

 \def\thebaroffset{0.0em}
}
{

 \def\Pmu         {\ensuremath{\mu}\xspace}                 
 \def\Pnu         {\ensuremath{\nu}\xspace}                 
                  
 \def\Ppi         {\ensuremath{\pi}\xspace}

 \def\Ppsi        {\ensuremath{\psi}\xspace}                 
                  
 \mathchardef\PDelta="7101
 \mathchardef\PXi="7104
 \mathchardef\PLambda="7103
 \mathchardef\PSigma="7106
 \mathchardef\POmega="710A
 \mathchardef\PUpsilon="7107
 \mathchardef\PPi="7105
 \def\PB      {\ensuremath{B}\xspace}                 
 \def\PD      {\ensuremath{D}\xspace}                 
 \def\PJ      {\ensuremath{J}\xspace}                 
 \def\PK      {\ensuremath{K}\xspace}                 
 \def\Pb      {\ensuremath{b}\xspace}                 
 \def\Pc      {\ensuremath{c}\xspace}                 
                  
 \def\Pe      {\ensuremath{e}\xspace}                 
 \def\Pp      {\ensuremath{p}\xspace}                 

 \def\Ps      {\ensuremath{s}\xspace}

 \def\thebaroffset{0.18em}
}
\newcommand{\offsetoverline}[2][\thebaroffset]{\kern #1\overline{\kern -#1 #2}}%

\makeatletter
\ifcase \@ptsize \relax
  \newcommand{\miniscule}{\@setfontsize\miniscule{4}{5}}
\or
  \newcommand{\miniscule}{\@setfontsize\miniscule{5}{6}}
\or
  \newcommand{\miniscule}{\@setfontsize\miniscule{5}{6}}
\fi
\makeatother

\DeclareRobustCommand{\optbar}[1]{\shortstack{{\miniscule (\rule[.5ex]{1.25em}{.18mm})}
  \\ [-.7ex] $#1$}}

\def\en         {{\ensuremath{\Pe^-}}\xspace}   
\def\ep         {{\ensuremath{\Pe^+}}\xspace}
\def\epm        {{\ensuremath{\Pe^\pm}}\xspace}

\def\mup        {{\ensuremath{\Pmu^+}}\xspace}
\def\mun        {{\ensuremath{\Pmu^-}}\xspace} 

\def\mumu       {{\ensuremath{\Pmu^+\Pmu^-}}\xspace}

\def\ellell     {\ensuremath{\ell^+ \ell^-}\xspace}

\def\neu        {{\ensuremath{\Pnu}}\xspace}

\def\squark    {{\ensuremath{\Ps}}\xspace}

\def\cquark    {{\ensuremath{\Pc}}\xspace}

\def\bquark    {{\ensuremath{\Pb}}\xspace}

\def\pion   {{\ensuremath{\Ppi}}\xspace}

\def\pip    {{\ensuremath{\pion^+}}\xspace}
\def\pim    {{\ensuremath{\pion^-}}\xspace}

\def\kaon    {{\ensuremath{\PK}}\xspace}

\def\KorKbar {\kern \thebaroffset\optbar{\kern -\thebaroffset \PK}{}\xspace}

\def\Kp      {{\ensuremath{\kaon^+}}\xspace}
\def\Km      {{\ensuremath{\kaon^-}}\xspace}
\def\Kpm     {{\ensuremath{\kaon^\pm}}\xspace}

\def\Kstarz  {{\ensuremath{\kaon^{*0}}}\xspace}

\def\Kstar   {{\ensuremath{\kaon^*}}\xspace}

\def\D       {{\ensuremath{\PD}}\xspace}

\def\DorDbar {\kern \thebaroffset\optbar{\kern -\thebaroffset \PD}\xspace}

\def\Dp      {{\ensuremath{\D^+}}\xspace}
\def\Dm      {{\ensuremath{\D^-}}\xspace}

\def\DpDm    {\ensuremath{\Dp {\kern -0.16em \Dm}}\xspace}

\def\Dsp     {{\ensuremath{\D^+_\squark}}\xspace}
\def\Dsm     {{\ensuremath{\D^-_\squark}}\xspace}

\def\B       {{\ensuremath{\PB}}\xspace}

\def\BorBbar {\kern \thebaroffset\optbar{\kern -\thebaroffset \PB}\xspace}
\def\Bz      {{\ensuremath{\B^0}}\xspace}

\def\Bd      {{\ensuremath{\B^0}}\xspace}

\def\BdorBdbar {\kern \thebaroffset\optbar{\kern -\thebaroffset \Bd}\xspace}

\def\Bs      {{\ensuremath{\B^0_\squark}}\xspace}

\def\BsorBsbar {\kern \thebaroffset\optbar{\kern -\thebaroffset \Bs}\xspace}

\def\jpsi     {{\ensuremath{{\PJ\mskip -3mu/\mskip -2mu\Ppsi}}}\xspace}
\def\psitwos  {{\ensuremath{\Ppsi{(2S)}}}\xspace}

\def\Y#1S{\ensuremath{\PUpsilon{(#1S)}}\xspace}

\def\proton      {{\ensuremath{\Pp}}\xspace}

\def\Lz          {{\ensuremath{\PLambda}}\xspace}

\def\LorLbar     {\kern \thebaroffset\optbar{\kern -\thebaroffset \PLambda}\xspace}

\def\Lb           {{\ensuremath{\Lz^0_\bquark}}\xspace}

\def\BF         {{\ensuremath{\mathcal{B}}}\xspace}

\newcommand{\decay}[2]{\mbox{\ensuremath{#1\!\to #2}}\xspace} 

\def\to                 {\ensuremath{\rightarrow}\xspace}

\def\qsq       {{\ensuremath{q^2}}\xspace}

\def\AT#1     {\ensuremath{A_{\mathrm{T}}^{#1}}\xspace}           

\def\C#1      {\ensuremath{\mathcal{C}_{#1}}\xspace}                       
\def\Cp#1     {\ensuremath{\mathcal{C}_{#1}^{'}}\xspace}                    
\def\Ceff#1   {\ensuremath{\mathcal{C}_{#1}^{\mathrm{(eff)}}}\xspace}        
\def\Cpeff#1  {\ensuremath{\mathcal{C}_{#1}^{'\mathrm{(eff)}}}\xspace}       
\def\Ope#1    {\ensuremath{\mathcal{O}_{#1}}\xspace}                       
\def\Opep#1   {\ensuremath{\mathcal{O}_{#1}^{'}}\xspace}                    

\newcommand{\aunit}[1]{\ensuremath{\text{\,#1}}}       

\newcommand{\tev}{\aunit{Te\kern -0.1em V}\xspace}
\newcommand{\gev}{\aunit{Ge\kern -0.1em V}\xspace}
\newcommand{\mev}{\aunit{Me\kern -0.1em V}\xspace}
\newcommand{\kev}{\aunit{ke\kern -0.1em V}\xspace}
\newcommand{\ev}{\aunit{e\kern -0.1em V}\xspace}
 
\newcommand{\mevc}{\ensuremath{\aunit{Me\kern -0.1em V\!/}c}\xspace}
\newcommand{\gevc}{\ensuremath{\aunit{Ge\kern -0.1em V\!/}c}\xspace}
\newcommand{\mevcc}{\ensuremath{\aunit{Me\kern -0.1em V\!/}c^2}\xspace}
\newcommand{\gevcc}{\ensuremath{\aunit{Ge\kern -0.1em V\!/}c^2}\xspace}
\newcommand{\gevgevcccc}{\ensuremath{\gev^2\!/c^4}\xspace} 

\def\fb   {\ensuremath{\aunit{fb}}\xspace}
\def\invfb   {\ensuremath{\fb^{-1}}\xspace}

\def\deriv {\ensuremath{\mathrm{d}}}

\def\gsim{{~\raise.15em\hbox{$>$}\kern-.85em
          \lower.35em\hbox{$\sim$}~}\xspace}
\def\lsim{{~\raise.15em\hbox{$<$}\kern-.85em
          \lower.35em\hbox{$\sim$}~}\xspace}

\def\pt         {\ensuremath{p_{\mathrm{T}}}\xspace}

\def\et         {\ensuremath{E_{\mathrm{T}}}\xspace}

\def\evtgen     {\mbox{\textsc{EvtGen}}\xspace}

\def\geant      {\mbox{\textsc{Geant4}}\xspace}

\def\photos     {\mbox{\textsc{Photos}}\xspace}

\def\pythia     {\mbox{\textsc{Pythia}}\xspace}

\def\tell1  {TELL1\xspace}
\def\ukl1   {UKL1\xspace}

\newcommand{\lhcborcid}[1]{\href{https://orcid.org/#1}{\hspace*{0.1em}\raisebox{-0.45ex}{\includegraphics[width=1em]{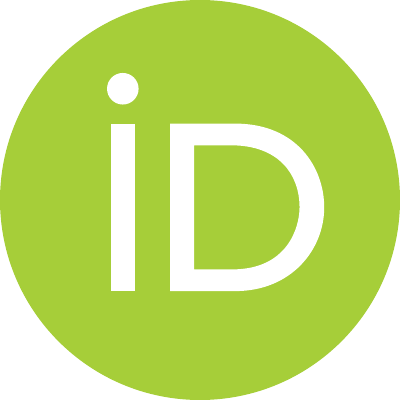}}}}


%% file: title-LHCb-PAPER.tex
\begin{titlepage}
\pagenumbering{roman}

\vspace*{-1.5cm}
\centerline{\large EUROPEAN ORGANIZATION FOR NUCLEAR RESEARCH (CERN)}
\vspace*{1.5cm}
\noindent
\begin{tabular*}{\linewidth}{lc@{\extracolsep{\fill}}r@{\extracolsep{0pt}}}
\ifthenelse{\boolean{pdflatex}}
{\vspace*{-1.5cm}\mbox{\!\!\!\includegraphics[width=.14\textwidth]{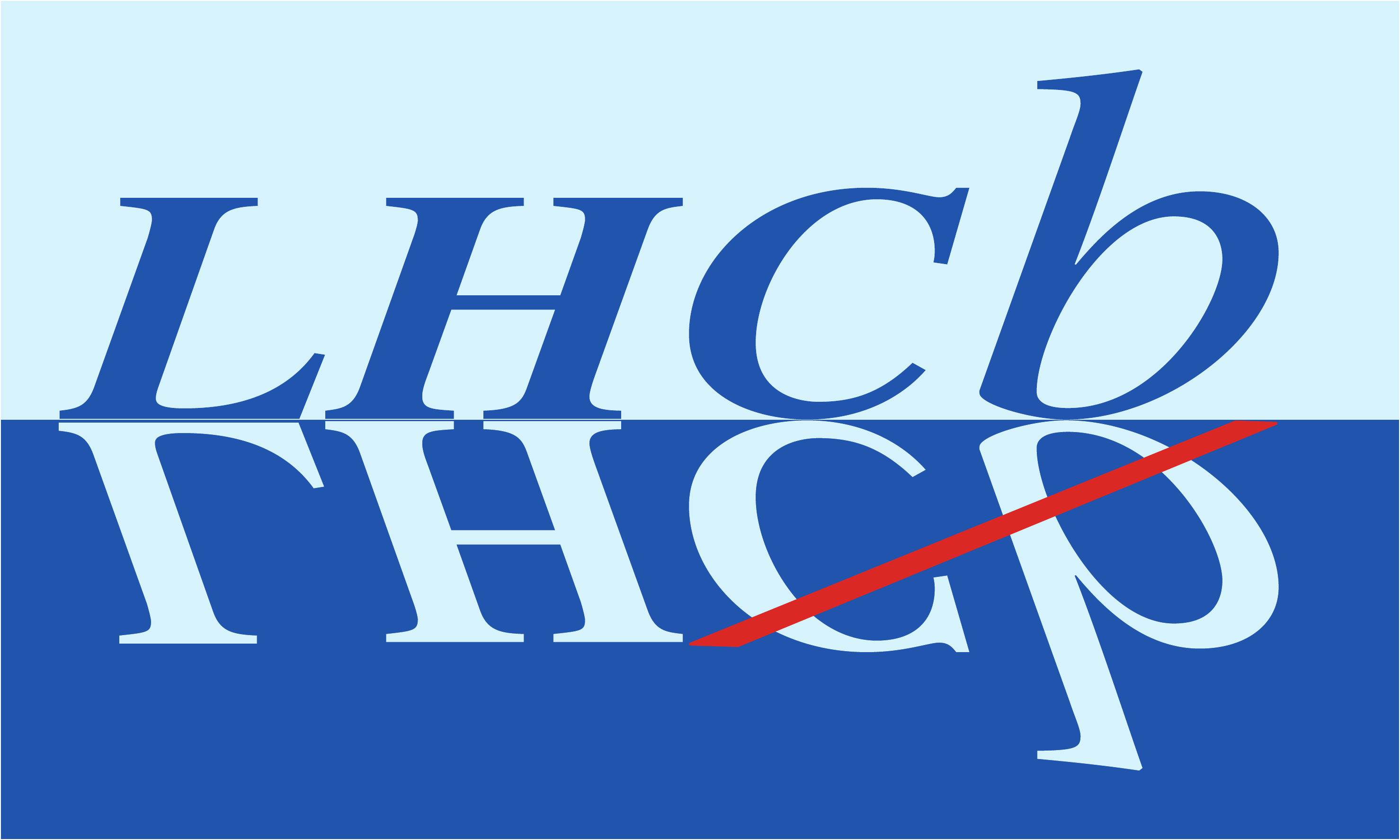}} & &}%
{\vspace*{-1.2cm}\mbox{\!\!\!\includegraphics[width=.12\textwidth]{figs/lhcb-logo.eps}} & &}%
\\
 & & CERN-EP-2024-255 \\  
 & & LHCb-PAPER-2024-032 \\  
 & & 7 April 2024 \\ 
 & & \\
\end{tabular*}

\vspace*{4.0cm}

{\normalfont\bfseries\boldmath\huge
\begin{center}
  \papertitle 
\end{center}
}

\vspace*{2.0cm}

\begin{center}
\paperauthors\footnote{Authors are listed at the end of this letter.}
\end{center}

\vspace{\fill}
\begin{abstract}
\noindent
Lepton flavour universality in rare $b\rightarrow s$ transitions is tested for the first time using $B_s^0$ meson decays. 
The measurements are performed using $pp$ collision data collected by the LHCb experiment between 2011 and 2018, corresponding to a total integrated luminosity of 9$\,{\rm fb}^{-1}$.  
Branching fraction ratios between the $B_s^0 \rightarrow \phi e^+e^-$ and $B_s^0 \rightarrow \phi \mu^+\mu^-$ decays are measured in three regions of dilepton mass squared, $q^2$, with $0.1 < q^2 < 1.1$, $1.1 < q^2 < 6.0$, and $15 < q^2 < 19\,{\rm GeV}^2/c^4$. 
The results agree with the Standard Model expectation of lepton flavour universality. 
\end{abstract}

\vspace*{2.0cm}

\begin{center}
  Published in Physical~Review~Letters \textbf{134} (2025), 121803
\end{center}

\vspace{\fill}

{\footnotesize 
\centerline{\copyright~\papercopyright. \href{\paperlicenceurl}{\paperlicence}.}}
\vspace*{2mm}

\end{titlepage}

\newpage
\setcounter{page}{2}
\mbox{~}

%% file: body.tex
The decay of a \Bs meson to a $\phi$ meson and pair of oppositely charged leptons, $\ellell$, involves a \bquark- to \squark-quark flavour-changing neutral-current transition. 
Such processes are rare in the Standard Model of particle physics (SM). 
In extensions of the SM, the rates of these processes could be modified by the contribution of as-yet undiscovered particles. 
The branching fraction of the \BsPhiMM decay, $\BF(\BsPhiMM)$, has been measured by the CDF~\cite{CDF:2011buy} and LHCb~\cite{LHCb-PAPER-2013-017,LHCb-PAPER-2015-023,LHCb-PAPER-2021-014} collaborations as a function of the dilepton mass squared, \qsq. 
The experimental measurements are systematically below theoretical predictions, with a local tension as large as 3.6 standard deviations~\cite{LHCb-PAPER-2021-014}. 
However, comparisons with theoretical predictions are complicated by sizeable uncertainties on the form factors for the $\Bs \to \phi$ transition~\cite{Horgan:2013hoa,Bharucha:2015bzk} and hadronic corrections to the decay rate~\cite{Ciuchini:2015qxb,Gubernari:2022hxn}. 
Discrepancies with SM predictions have also been found in the angular distributions of the \BsPhiMM decay~\cite{LHCb-PAPER-2021-022}, as well as in the rate and angular distribution of other \decay{\bquark}{\squark\mumu} transitions~\cite{LHCb-PAPER-2014-006,CMS:2017rzx,ATLAS:2018gqc,LHCb-PAPER-2020-002,LHCb-PAPER-2020-041,LHCb-PAPER-2023-033,LHCb-PAPER-2024-011,CMS:2024syx,CMS:2024atz}. 
This pattern of tensions with SM predictions has led to significant discussion within the community~\cite{Crivellin:2023zui,Capdevila:2023yhq,Athron:2023hmz,Allanach:2023uxz,Davighi:2023evx,Isidori:2023unk,Ciuchini:2022wbq,Greljo:2022jac,SinghChundawat:2022zdf,Parrott:2022zte,FernandezNavarro:2022gst,SinghChundawat:2022ldm,Allanach:2022iod,Bause:2022rrs, Ciuchini:2021smi,Alguero:2023jeh}.

Lepton flavour universality tests, which compare the rates of processes involving different flavours of charged leptons, can provide powerful probes of the SM. 
In such comparisons, SM uncertainties associated with hadronic form factors largely cancel and, since the electroweak couplings are flavour-universal, processes are expected to differ only by phase-space and QED effects~\cite{Bordone:2016gaq,Isidori:2022bzw}. 
In some extensions of the SM, this accidental flavour-symmetry could be violated (see for example Refs.~\cite{Hiller:2003js,Glashow:2014iga,Dorsner:2016wpm,Buttazzo:2017ixm}). 

In this Letter, a first measurement of the lepton universality ratio $\rphi = \BF(\BsPhiMM)/\BF(\BsPhiEE)$ is reported.
Inclusion of charge-conjugate processes is implied throughout.  
The analysis uses data from $pp$ collisions at centre-of-mass energies of $\sqrt{s} = 7$, $8$, and $13 \tev$, corresponding to a total integrated luminosity of 9\invfb, collected by the LHCb collaboration between 2011 and 2018. 
Measurements are performed in three bins of \qsq: $0.1 < \qsq < 1.1$ (low), $1.1 < \qsq < 6.0$ (central), and $15 < \qsq < 19\gevgevcccc$ (high). 
The lowest \qsq bin contains small contributions from light-quark resonances decaying to \ellell (predominantly the $\rho(770)^{0}$ and $\phi$ mesons).
Similar tests involving \decay{B^{\{+,0\}}}{K^{\{+,0\}}\ellell}, \decay{B^{\{+,0\}}}{\Kstar^{\{+,0\}}\ellell} and \decay{\Lb}{p\Km\ellell} decays have previously been performed by the \babar~\cite{BaBar:2012mrf}, \belle~\cite{Belle:2019xld}, \cms~\cite{CMS:2024syx} and LHCb collaborations~\cite{LHCb-PAPER-2014-024,LHCb-PAPER-2017-013,LHCb-PAPER-2019-040,LHCb-PAPER-2021-004,LHCb-PAPER-2021-038,LHCb-PAPER-2022-045,LHCb-PAPER-2022-046}. 
One advantage of performing tests with \BsPhiLL decays is that the clean signature of the \decay{\phi}{\Kp\Km} decay significantly reduces most sources of background. 
This facilitates the inclusion of the high-\qsq bin, which has not been studied in previous LHCb publications.

The analysis strategy follows closely the approach described in Refs.~\cite{LHCb-PAPER-2022-045,LHCb-PAPER-2022-046}. 
To reduce systematic uncertainties on the modelling of the lepton reconstruction efficiency, the double ratio of branching fractions, 
\begin{equation}
\rphi =  
\left(\frac{\BF(\BsPhiMM)}{\BF(\decay{\Bs}{\jpsi(\to \mumu) \phi})}\right) \Big/\left(\frac{\BF(\BsPhiEE)}{\BF(\decay{\Bs}{\jpsi(\to \ep\en) \phi})}\right)\,, 
\label{eq:rphi-def}
\end{equation}
is measured. 
Here, each individual branching fraction is obtained as the corresponding yield in the relevant \qsq\ interval divided by its efficiency, with other multiplicative factors that cancel exactly in Eq.~\eqref{eq:rphi-def}.  
In practice, \invrphi is measured rather than \rphi such that the small \decay{\Bs}{\phi\ep\en} yield appears in the numerator and the statistical behaviour of the observable more closely follows a Gaussian distribution. 
To avoid experimenter's bias, the \BsPhiEE candidates with masses in the range 5.2--5.5\gevcc (slightly different for each \qsq interval), the corresponding signal yields, and \invrphi are not inspected until all aspects of the analysis procedure are finalised.

The \lhcb detector~\cite{LHCb-DP-2008-001,LHCb-DP-2014-002} is a single-arm forward spectrometer covering the \mbox{pseudorapidity} range $2<\eta <5$, designed for the study of particles containing \bquark or \cquark
quarks. 
The detector includes a high-precision tracking system, comprising silicon-strip~\cite{LHCb-DP-2014-001, LHCb-DP-2013-003} and straw drift-tube~\cite{LHCb-DP-2017-001} detectors. 
Different types of charged hadrons are distinguished using information from two ring-imaging Cherenkov detectors~\cite{LHCb-DP-2012-003}.
Electrons are identified using a calorimeter system comprising scintillating-pad and preshower detectors, an electromagnetic, and a hadronic calorimeter.
Muons are identified by a system composed of alternating layers of iron and multiwire proportional chambers~\cite{LHCb-DP-2012-002}.
The online event selection is performed by a trigger~\cite{LHCb-DP-2012-004}, which consists of a hardware and a software stage. 

Simulated data samples are used to determine the efficiency of the event reconstruction and candidate selection, which is needed to determine the branching fractions from the observed yields. 
These samples are also used to understand candidate mass lineshapes, and the contributions from different background sources. 
In the simulation, $pp$ collisions are generated using \pythia~\cite{Sjostrand:2007gs,*Sjostrand:2006za} with a specific \lhcb configuration~\cite{LHCb-PROC-2010-056}.
Decays of unstable particles are described by \evtgen~\cite{Lange:2001uf}, in which final-state radiation is generated using \photos~\cite{davidson2015photos}.
Simulated \BsPhiLL decays are generated according to a model based on Refs.~\cite{Ali:1999mm,Descotes-Genon:2015hea,Ball:2004rg}. 
The interactions of the generated particles with the detector, and its response, are simulated using the \geant toolkit~\cite{Allison:2006ve, *Agostinelli:2002hh} as described in Ref.~\cite{LHCb-PROC-2011-006}.
The simulated samples are corrected to account for known data-simulation differences in the \Bs production kinematics and detector occupancy, as well as track reconstruction, particle identification and trigger efficiencies. 
The \qsq distribution of the simulated \BsPhiEE  sample is smeared to better match the resolution of the \BsJpsiPhi candidates in data. 

The analysis uses data triggered at the hardware stage either by particles in the event that are not used to form signal candidates, or by an electron (muon) with large transverse energy (transverse momentum) for \BsPhiEE (\BsPhiMM) decays. 
The \et (\pt) thresholds vary in the ranges 2.4--3.0\gev (1.5--2.9\gevc).
In the subsequent software trigger, events are retained if they contain at least one reconstructed track that has a large \pt and is significantly displaced from every primary proton-proton interaction vertex (PV). 
This track is then combined with at least one other track to form a vertex and events are retained based on topological criteria~\cite{BBDT,LHCb-PROC-2015-018}. 
Offline, \BsPhiLL candidates are formed by combining two leptons of opposite charge with a pair of kaons of opposite charge. 
A dedicated algorithm is used to correct for the bremsstrahlung emission of the electrons, associating energy deposits in the calorimeter to electrons based on their trajectories.  
Candidates are accepted if their reconstructed mass is in the range $5000<\mKKMM<6000$ or $4600<\mKKEE<6600\mevcc$ and if the kaon pair has a mass within $\pm 12\mevcc$ of the known $\phi$ meson mass~\cite{PDG2024}.  
Contributions in this mass region from \decay{\Bs}{\Kp\Km\ellell} decays, where the kaons do not originate from a $\phi$ meson, are known to be small and are neglected~\cite{LHCb-PAPER-2021-014}. 
The leptons and kaons must have significant impact parameter~(IP) values and form a common vertex that is displaced from every PV. 
The PV that fits best to the \Bs flight direction is taken as the associated PV, and the \Bs candidate must have a small IP with respect to this PV.
In the \BsPhiMM sample, candidates with dimuon masses within $\pm 100\mevcc$ of the known \jpsi mass are assumed to originate from \BsJpsiPhi decays. 
A wider range, $6 < \qsq < 11\gevgevcccc$, is used for \BsJpsiPhiEE decays to account for the worse mass resolution resulting from bremsstrahlung and the associated energy recovery procedure. 

Combinatorial backgrounds are suppressed using a gradient boosted decision tree~(BDT) classifier~\cite{CatBoost}. 
The classifier uses topological and kinematic information, including: the chi-square of a common vertex fit to the four particles, the consistency of the \Bs candidate with originating from the associated PV, the displacement of the candidate's decay vertex from that PV, its \pt, and the \pt and IP of the kaons and the leptons. 
Separate classifiers are trained for the dielectron and dimuon final states and for the 2011--2012, 2015--2016 and 2017--2018 run periods. 
Simulated signal decays and candidates from upper mass sidebands, with $\mKKMM > 5400$ and $\mKKEE > 5600\mevcc$, are used as signal and background proxies in the training. 
To increase the amount of data available for training and avoid biases, $k$-fold cross-validation~\cite{kfold} with $k=10$ is used for the training/testing samples. 
The working points of the classifier are chosen to optimise the precision on an $R_{\phi}$ measurement based on the measured \BsPhiMM branching fraction~\cite{LHCb-PAPER-2021-014}, assuming $\rphi=1$. 
Working points are optimised separately for the three \qsq regions. 

The narrow window around the $\phi$ meson mass reduces the background from many specific \bquark-hadron decays.
Backgrounds from \decay{\Bz}{\Kstarz(\to \Kp\pim)\ellell} and \decay{\Lb}{\proton\Km\ellell} decays are reduced to a negligible level by particle identification requirements.  
Potential background from \decay{\Bs}{\Dsm(\to\phi\pim)\ell^{+}\neu_{\ell}} decays, where the \pim is mistakenly identified as a lepton, is rejected by removing candidates if the $\phi \ell^-$ mass is consistent with that of a \Dsm meson after  the pion mass  has been assigned to the lepton. 
This veto also removes potential background from \decay{\Bs}{\Dsm(\to\phi\pim)\pip} decays with double misidentification.
Additional particle identification criteria are applied to remove \bquark-hadron decays to \jpsi or \psitwos mesons and two charged hadrons, where a hadron is mistakenly reconstructed as a lepton and vice versa.  
These criteria are applied for \BsPhiMM candidates if the mass formed by the hadron and muon of opposite charge is consistent with the \jpsi or \psitwos mass when the hadron is assigned the muon mass. 
For \BsPhiEE candidates the criteria are applied if the four-body mass, calculated with the electron mass assigned to the \Kpm, and the $e^{\pm}$ assigned either the pion, kaon or proton mass, is close to that of a known \bquark-hadron.

The $\phi\ellell$ final state has relatively few potential backgrounds from partially reconstructed $\bquark$-hadron decays with missing hadrons. 
The most relevant are \decay{\B^{\{0,+\}}}{\phi K^{\{0,+\}}\ellell} decays where the kaon is not reconstructed. 
Backgrounds from hadronic $\bquark$ decays, where two charged hadrons are misidentified as leptons and potentially one or more hadrons are not reconstructed, are poorly known.
The contribution from these backgrounds is estimated using the same data-driven approach as employed in Refs.~\cite{LHCb-PAPER-2022-045,LHCb-PAPER-2022-046}.

A separate BDT classifier is trained to reject background in the high-\qsq interval from \decay{\jpsi}{\ep\en} decays originating from a \bquark-hadron, where bremsstrahlung photons have been incorrectly assigned to the electrons, or incorrectly reconstructed.  
The BDT classifier is trained using simulated events and uses information on the bremsstrahlung photon energy, reconstructed track momentum, track asymmetries and the dielectron mass. 
The classifier achieves a signal efficiency of $90\%$ while rejecting more than $80\%$ of these backgrounds.

After applying the full selection less than $0.3\%$ of events contain more than one candidate, in which case a single candidate is selected at random.
The yields of \BsPhiLL and \BsJpsiPhi decays are determined by performing unbinned extended maximum-likelihood fits to the mass distributions of the candidates in the different samples, using \zfit~\cite{zfit}. 
The fit is performed simultaneously, where \invrphi is shared over the different data-taking periods. 
The \BsPhiLL and \BsJpsiPhi signals are described by modified Gaussian functions with power-law tails. 
The functions have different width and tail parameters on the left- and right-hand sides of the distribution. 
For \BsPhiEE decays, the lineshape is obtained by summing functions that represent the cases where either zero, one, or two or more calorimeter clusters are associated with bremsstrahlung from the dielectron pair. 
The fractional contribution of these functions is validated to agree well between data and simulation using the \decay{\Bs}{\jpsi\phi} sample.
These fractions are then fixed from simulation, together with the tail parameters of the signal shapes, in the fit to data. 
The peak and width parameters for each \qsq bin are taken from simulation but are allowed to vary by scale factors in the fits to the data. 
For the \BsPhiLL samples, the scale factors are constrained to match those of the corresponding \BsJpsiPhi samples. 

Combinatorial background in the low- and central-\qsq bins is described by an exponential function using independent slope parameters in the different \qsq-regions. 
The combinatorial shape is modified at high \qsq, by multiplying the exponential function by a sigmoid function to account for the reduced \qsq range available at low \mKKLL. 
The shape of the sigmoid is determined from a model where \mKKLL and \qsq are described by exponential functions, which are parameterised using candidates at large \mKKLL. 
It is verified with data that there is negligible correlation between the \qsq  and \mKKLL distributions.

Background from semileptonic decays of a \Bs to a \Dsm meson, where the \Dsm also decays semileptonically, is modelled by an exponential function with a slope parameter fixed from simulation. 
This background is only present in the low- and central-\qsq bins. 
The shape and level of background from misidentified hadrons are obtained by combining three control regions in data, where either one or both leptons are selected with particle identification criteria favouring hadrons, following the procedure in Refs.~\cite{LHCb-PAPER-2022-045,LHCb-PAPER-2022-046}. 

For the fit to the dielectron samples, additional background components, with parameters fixed from simulation, are included to account for leakage from \BsJpsiPhi (\BsPsiTwoSPhi) decays into the central- and high-\qsq (high-\qsq only) bins. 
In central \qsq, the \BsJpsiPhi leakage is described by a Gaussian function.
At high \qsq, the \BsJpsiPhi and \BsPsiTwoSPhi leakage shapes are described by a nonparametric kernel density estimate and by a Johnson $S_U$ function~\cite{JohnsonSU}, respectively. 
For the \BsJpsiPhi samples, background components below the $1\%$ level are included to describe \decay{\Bz}{\jpsi\Kstarz}, \decay{\Lb}{\jpsi p\Km} and \decay{\Bz}{\jpsi\phi} decays. 
These components are taken nonparametrically from simulated samples. 
The yields of semileptonic, leakage and misidentified backgrounds are each constrained in the fits to estimates from simulation.  
 
\begin{figure}[!tb] 
    \centering
    \includegraphics[width=\textwidth]{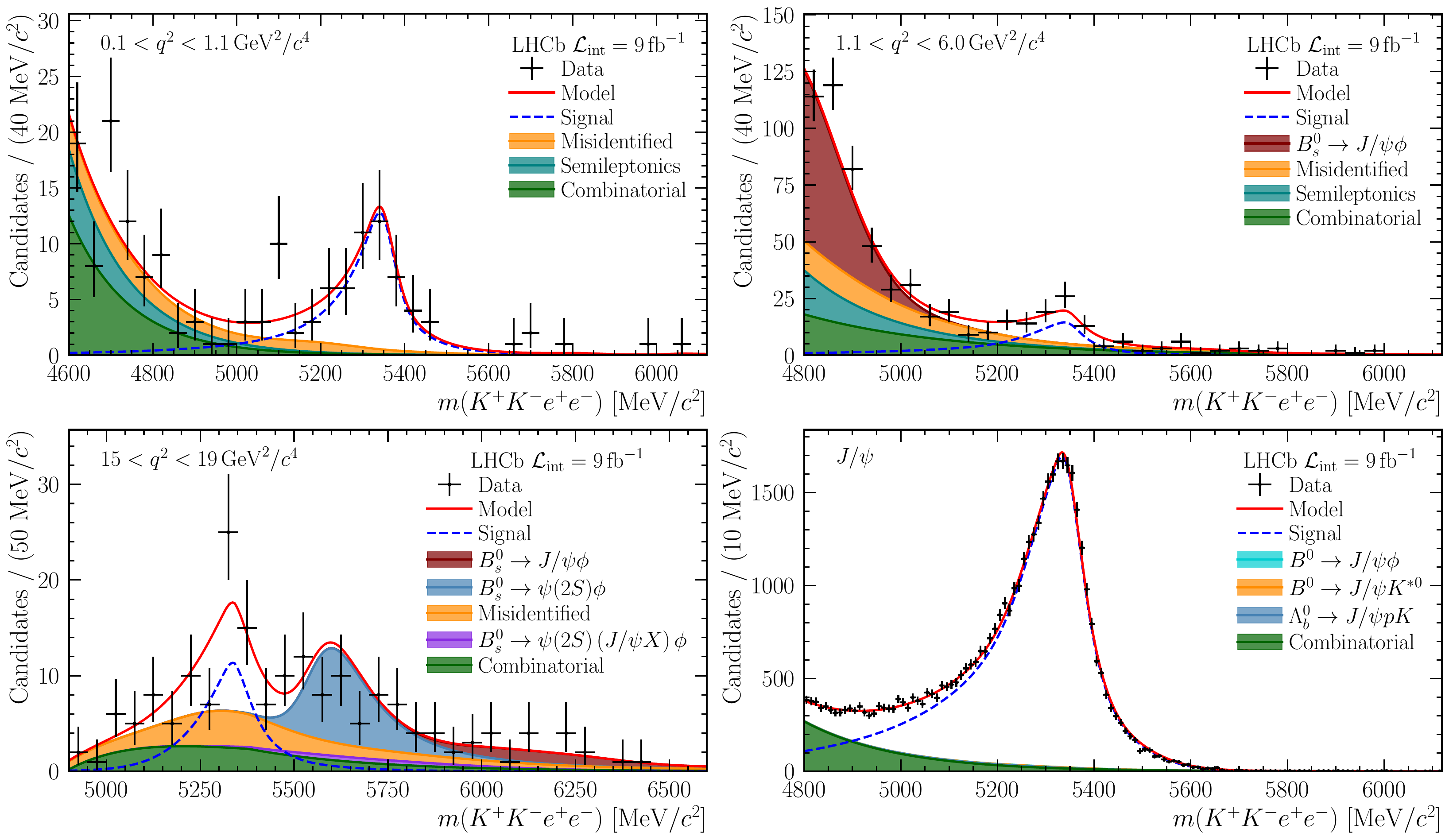}
    \caption{
    Mass distributions of selected $B_s^0 \rightarrow \phi e^+ e^-$ candidates in the (top left) low-, (top right) central-, (bottom left) high-$q^2$, and (bottom right) $J/\psi$ bins. 
    The data are summed over the three data-taking periods and are compared with the result of the fit described in the text. 
    }
    \label{fig:phill}
\end{figure}

Figure~\ref{fig:phill} shows the mass distributions of selected \BsPhiEE candidates in the different \qsq bins. 
The corresponding mass distributions for selected \BsPhiMM candidates, which are similar to those in Ref.~\cite{LHCb-PAPER-2021-014}, are provided as Appendixes. 
The yields of the \BsPhiEE (\BsPhiMM) decays in the low-, central-, and high-\qsq bin are $63\pm 9$ ($159\pm 13$), $119\pm 14$ ($300\pm 18$) and $50\pm 13$ ($312\pm 18$), respectively. 
The lower yield for the dielectron- compared to the dimuon-modes is due to lower track-reconstruction efficiency for electrons and the large \et requirement necessary to select them in the hardware trigger. 
The statistical significance of the \BsPhiEE signal in the three \qsq bins is determined using Wilks' theorem~\cite{Wilks:1938dza} to correspond to 6.8, 5.4 and 3.6 standard deviations ($\sigma$).

The analysis is validated using pseudoexperiments, generated with $\rphi=1$ and taking $\BF(\BsPhiMM)$ from Ref.~\cite{LHCb-PAPER-2021-014}. 
The resulting fits are found to have good coverage 
and bias at the level of 10\% of the statistical uncertainty or less.
The bias is corrected and the uncertainty on the bias assigned as a source of systematic uncertainty.  
Several cross-checks are performed to validate the analysis using data. 
The branching fraction of the \BsPhiMM decay is measured and found to be consistent with Ref.~\cite{LHCb-PAPER-2021-014}.
Consistent values of \invrphi are found when analysing the data with stricter particle identification requirements. 
The \BsJpsiPhi yields are also determined from fits to the candidate mass distribution with the \ellell mass constrained to the known mass of the \jpsi meson. 
This significantly improves the mass resolution on the \BsJpsiPhi samples but only changes the yields by 0.2\%, confirming that the $\Kp\Km\ellell$ mass model gives an appropriate description of the data. 
The ratio of branching fractions between \BsJpsiPhiMM and \BsJpsiPhiEE decays, \rjpsi, is determined from fits to the constrained masses, in bins of different  kinematic variables, and found to be consistent with unity; 
the most significant variation is used to assign a systematic uncertainty on the measurement  of \invrphi. 
The value of \rjpsi for the combined data set is $\rjpsi = 0.997\pm 0.013$, where the uncertainty combines the statistical and a subset of the systematic uncertainties pertaining to the limited size of the calibration samples used in the data-simulation corrections. 
The double ratio \Rpsitwos, defined by replacing the signal mode \BsPhiLL\ with the decay \BsPsiTwoSPhi\ in Eq.~\ref{eq:rphi-def}, is determined from fits to candidates with $11 < \qsq < 15\gevgevcccc$.
The double ratio is found to be $\Rpsitwos = 1.010 \pm 0.026$, consistent with the expectation of unity, where again the uncertainty combines statistical and systematic contributions. 
An independent test~\cite{LHCb-PAPER-2023-038} using $\phi$ mesons from \Dsp decays, that decay to $\ep\en$ and $\mumu$, shows that the lepton efficiency is well understood even for lower momentum leptons.

The different sources of systematic uncertainty on \invrphi considered in the analysis are summarised in Table~\ref{table:systematics}. 
Sources of uncertainty on the efficiencies, related to the finite size of the simulation and variations of the data-simulation corrections, impact \invrphi at the $\sim 1\%$ level. 
Small variations of the decay model used in the simulation, accounting for the time dependence of the signal decay rate, have a minimal impact on \invrphi. 
Variations of the \qsq resolution also impact the efficiency calculation as simulated decays migrate into and out of the \qsq bins used in the analysis.
A systematic uncertainty on the \qsq-resolution correction is determined by evaluating separate correction factors from the \BsJpsiPhi sample in bins of electron \pt.  
Modifying the correction factors changes \invrphi by less than 1\%. 

The largest sources of systematic uncertainty are associated with the modelling of the signal and background in the likelihood fits. 
The uncertainty associated with the signal lineshape is estimated by replacing the model by a simplified function, with a single Gaussian core, and by varying the fractions of the different bremsstrahlung categories that form the lineshape. 
The uncertainty on the modelling of misidentified hadronic backgrounds is determined by varying both the size and shape of the contribution in the fit. 
Variations are made to the particle-identification requirements used to define the background-like control regions, the efficiency maps that translate yields in the background-like regions to yields in the signal region, and in the model used to  parameterise the lineshape. 
The effect of the finite size of the background-like samples is assessed using a bootstrapping technique~\cite{efron:1979}. 
Uncertainty on the high-\qsq combinatorial-background shape is evaluated by varying the shape of the $\qsq$ distribution used to determine the sigmoid parameters, and by replacing the model by an alternate function with a power-law turn-on and exponential fall-off, with parameters constrained from a fit to data with same-sign $\ell^{\pm}\ell^{\pm}$ or $\kaon^{\pm}\kaon^{\pm}$ combinations. 
An uncertainty from the \jpsi and \psitwos leakage into the central- and high-\qsq bins is assigned by replacing the parametric shapes by a nonparametric model, varying the constraints on the background yields in the fit, and by varying the \qsq resolution in the simulation. 
An uncertainty associated with the semileptonic background is evaluated by estimating the effects of replacing the background model by a nonparametric model and varying the background level.

\begin{table}[!tb]
  \caption{
  Sources of systematic uncertainty in $10^{-2}$ on the $R_\phi^{-1}$ measurement in the low-, central- and high-$q^2$ bins. 
  The total uncertainty is computed by combining the contributions from individual sources in quadrature. 
  }
  \centering
  \label{table:systematics} 
  \begin{tabular}{lrcc}
      \toprule
      Source $[10^{-2}]$       & low    & central & high \\ 
      \midrule
      Fit bias                 & $<$0.1 & 0.9     & 1.4 \\ 
      Normalisation            & 0.2    & 0.2     & 0.2 \\ 
      \rjpsi variation         & 1.3    & 0.6     & 0.9 \\
      Efficiency calibration   & 0.7    & 0.5     & 1.2 \\ 
      \qsq smearing            & 0.6    & 0.5     & 0.3 \\ 
      Decay model              & 0.1    & 0.3     & 0.1 \\ 
      Signal lineshape         & 1.7    & 1.5     & 3.7 \\
      Misidentified background & 4.1    & 4.3     & 7.4 \\ 
      Combinatorial background & --~    & --      & 3.9 \\ 
      Leakage background       & --~    & 0.9     & 2.1 \\
      Semileptonic background  & 1.3    & 1.2     & -- \\
      \midrule
      Total                    & 4.9    & 4.9     & 9.6  \\
      \bottomrule
  \end{tabular}
\end{table} 

\begin{figure}[!tb]
    \begin{center} 
    \includegraphics[width=0.6\linewidth]{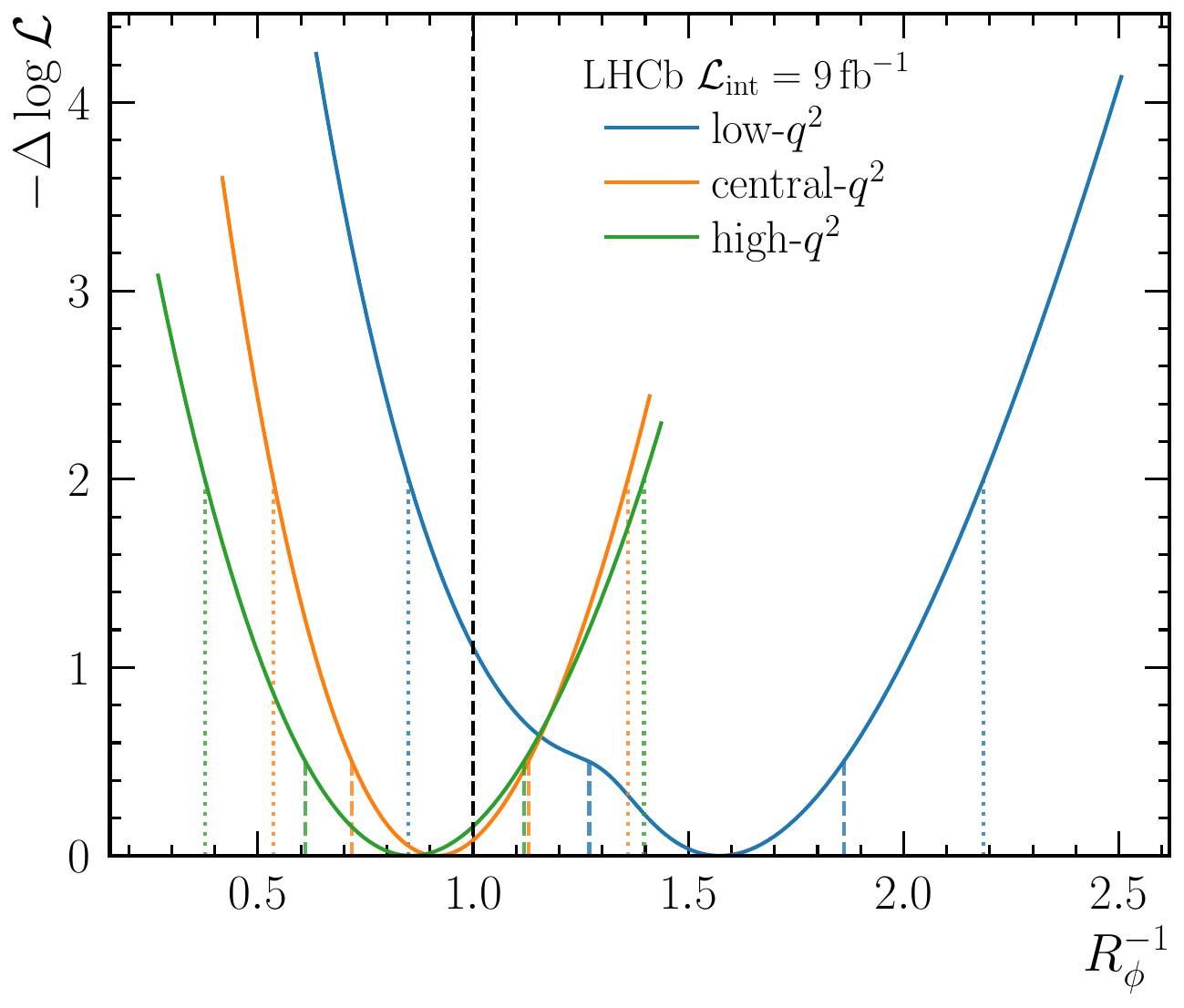} 
    \end{center} 
    \caption{
    Profile log-likelihood of $R_{\phi}^{-1}$ for the  low-, central- and high-$q^2$ bins, relative to the best fit point. 
    Systematic uncertainties are included by convolving the likelihood from the fit with a Gaussian distribution of width equal to the systematic uncertainty. 
    The vertical lines indicate the $1$ and $2\,\sigma$ confidence intervals, respectively. 
    }
    \label{fig:likelihood}
\end{figure}

The \invrphi values in the three \qsq ranges are given in Table~\ref{tab:results}. 
The results are in good agreement with the SM expectation. 
Figure~\ref{fig:likelihood} shows the variation with \invrphi of the difference in log-likelihood of the fit from the best fit point.  
For low-\qsq, a local minimum around $\invrphi \sim 1.3$ yields a nonparabolic profile likelihood towards lower \invrphi values. 
The presence of this additional minimum is evident when considering the variation of the likelihood in the full parameter space and is induced by large correlations between the combinatorial background and the signal model in combination with the small sample size. 
Depending on the \mKKEE mass range, the particle identification criteria and BDT requirement, the minimum at $\invrphi\sim 1.3$ can be favoured. 
The 2\,$\sigma$ interval on \invrphi from the profile likelihood is $[0.86, 2.18]$; the results is in agreement at the level of 1.4\,$\sigma$ with a SM prediction of 1.016 computed using Ref.~\cite{Straub:2018kue}.
Table~\ref{tab:results} also presents the differential branching fraction of the \BsPhiEE decay, determined from \invrphi,   
\mbox{$\deriv\BF(\decay{\Bs}{\phi\mumu})/\deriv\qsq/\BF(\decay{\Bs}{\jpsi\phi})$}
 ~\cite{LHCb-PAPER-2021-014} and \mbox{$\BF(\decay{\Bs}{\jpsi\phi})$}~\cite{LHCb-PAPER-2020-046,PDG2024}. 
The upper value of the low-\qsq range in Ref.~\cite{LHCb-PAPER-2021-014} is extrapolated from 0.98\gevgevcccc to 1.1\gevgevcccc by assuming that $\deriv\BF(\BsPhiMM)/\deriv\qsq$ is uniform in \qsq. 
At 2\,$\sigma$, $\deriv\BF(\BsPhiEE)/\deriv\qsq$ in the low-\qsq region is in the range  $[0.65, 2.08]\times 10^{-7}\invgevgevcccc$.
As implied by \invrphi, the \BsPhiEE branching fraction is consistent with that of the \BsPhiMM decay and lies below the central value of the SM prediction in the central- and high-\qsq bins. 

\begin{table}[!tb] 
\caption{ 
Values of \invrphi and $\deriv \mathcal{B}(B_s^0 \rightarrow \phi e^+ e^-)/\deriv q^2$ in the low-, central- and high-$q^2$ bins. 
The first uncertainty is statistical and the second systematic. 
For the differential branching fraction, the third and the fourth uncertainty are due to the experimental uncertainty on the ratio $\deriv\mathcal{B}(B_s^0 \rightarrow \phi\mu^+\mu^-)/\deriv q^2/\mathcal{B}(B_s^0 \rightarrow J/\psi \phi)$~\cite{LHCb-PAPER-2021-014} and on $\mathcal{B}(B_s^0 \rightarrow J/\psi \phi)$~\cite{LHCb-PAPER-2020-046,PDG2024}, respectively.
}
\centering
\begin{tabular}{ccc}
\toprule
\qsq $[\gevgevcccc]$ & \invrphi & $\deriv \BF(\BsPhiEE)/\deriv\qsq$ $[ 10^{-7}\invgevgevcccc]$\\ 
\midrule
$0.1 < \qsq < 1.1$ & $1.57\,^{+0.28}_{-0.25} \pm 0.05$ & $1.38\;\;\;{}^{+0.25}_{-0.22}\; \pm 0.04 \pm 0.19 \pm0.06$ \\ 
$1.1 < \qsq < 6.0$ & $0.91\,^{+0.20}_{-0.19} \pm 0.05$ & $0.26 \pm 0.06 \pm 0.01 \pm 0.01 \pm 0.01$ \\ 
$15.0 < \qsq < 19.0$ & $0.85\,^{+0.24}_{-0.23} \pm 0.10$ & $0.39 \pm 0.11 \pm 0.04 \pm 0.02 \pm 0.02$ \\
\bottomrule
\end{tabular} 
\label{tab:results}
\end{table} 
 
In summary, the first measurement of the ratio of \BsPhiEE and \BsPhiMM branching fractions is presented. 
This represents the first test of lepton flavour universality with \BsPhiLL decays.
It is also the most precise test of lepton flavour universality in the high-\qsq region of any measurement involving \decay{\bquark}{\squark\ellell} transitions. 
Together with Ref.~\cite{LHCb-PAPER-2024-030}, this result constitutes the first observation of the \BsPhiEE decay.  
In addition, the differential branching fraction of the \BsPhiEE decay is measured in three bins of \qsq. 
The results are statistically limited and agree with the SM expectation of lepton flavour universality, placing important constraints on extensions of the SM.

\clearpage

%% file: acknowledgements.tex
\section*{Acknowledgements}
%
%
\noindent We express our gratitude to our colleagues in the CERN
accelerator departments for the excellent performance of the LHC. We
thank the technical and administrative staff at the LHCb
institutes.
We acknowledge support from CERN and from the national agencies:
CAPES, CNPq, FAPERJ and FINEP (Brazil);
MOST and NSFC (China);
CNRS/IN2P3 (France);
BMBF, DFG and MPG (Germany);
INFN (Italy);
NWO (Netherlands);
MNiSW and NCN (Poland);
MCID/IFA (Romania);
MICIU and AEI (Spain);
SNSF and SER (Switzerland);
NASU (Ukraine);
STFC (United Kingdom);
DOE NP and NSF (USA).
We acknowledge the computing resources that are provided by CERN, IN2P3
(France), KIT and DESY (Germany), INFN (Italy), SURF (Netherlands),
PIC (Spain), GridPP (United Kingdom),
CSCS (Switzerland), IFIN-HH (Romania), CBPF (Brazil),
and Polish WLCG (Poland).
We are indebted to the communities behind the multiple open-source
software packages on which we depend.
Individual groups or members have received support from
ARC and ARDC (Australia);
Key Research Program of Frontier Sciences of CAS, CAS PIFI, CAS CCEPP,
Fundamental Research Funds for the Central Universities,
and Sci. \& Tech. Program of Guangzhou (China);
Minciencias (Colombia);
EPLANET, Marie Sk\l{}odowska-Curie Actions, ERC and NextGenerationEU (European Union);
A*MIDEX, ANR, IPhU and Labex P2IO, and R\'{e}gion Auvergne-Rh\^{o}ne-Alpes (France);
AvH Foundation (Germany);
ICSC (Italy);
Severo Ochoa and Mar\'ia de Maeztu Units of Excellence, GVA, XuntaGal, GENCAT, InTalent-Inditex and Prog. ~Atracci\'on Talento CM (Spain);
SRC (Sweden);
the Leverhulme Trust, the Royal Society
 and UKRI (United Kingdom).

%% file: appendix.tex

{\noindent\normalfont\bfseries\Large Appendices}

\appendix

\section{Mass fits for \texorpdfstring{\BsPhiMM}{Bs -> phi mu mu} decays}

\begin{figure}[!b]
    \centering
    \includegraphics[width=\linewidth]{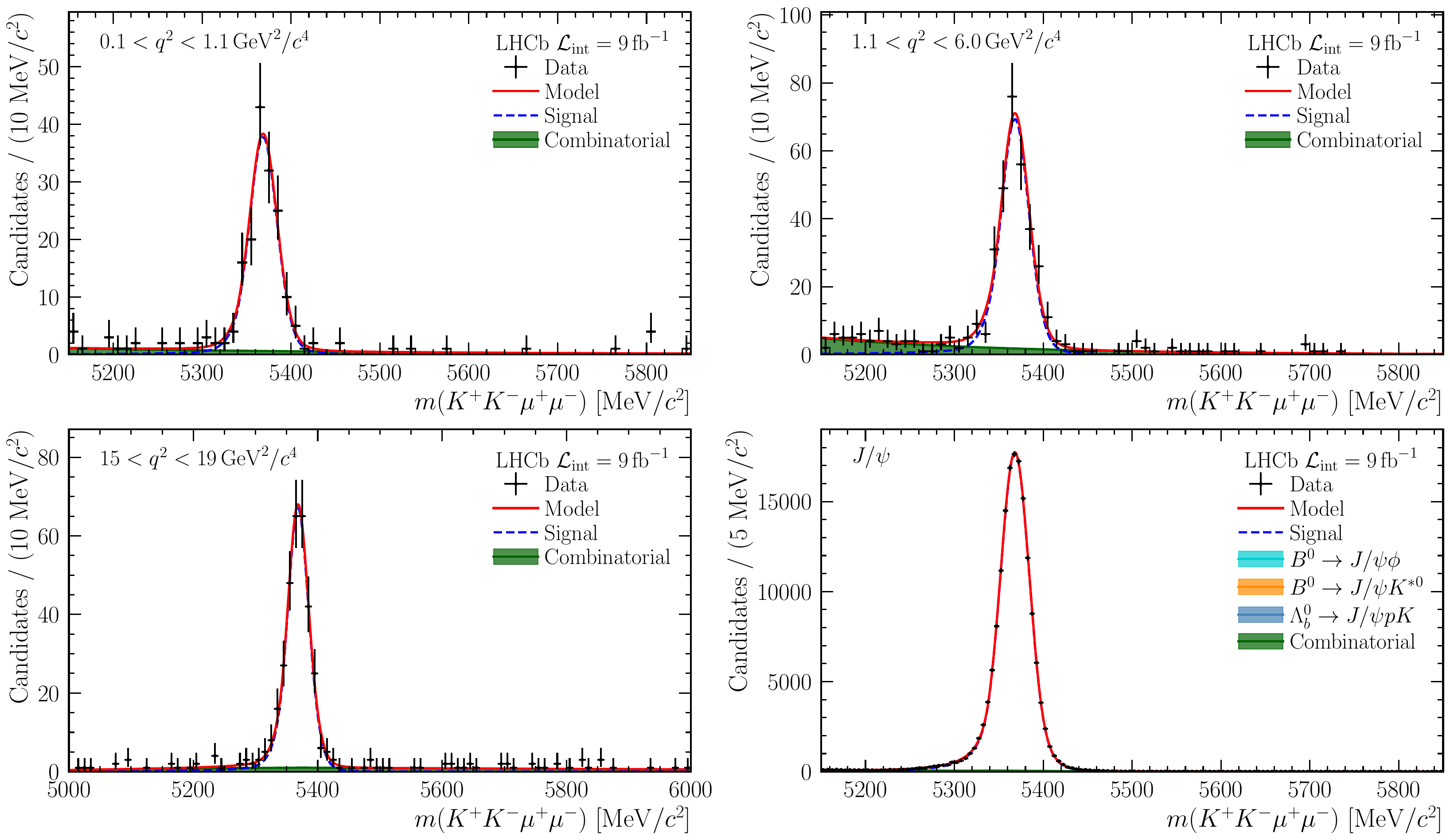}
    \caption{    
    Mass distributions of selected $B_s^0 \rightarrow \phi \mu^+ \mu^-$ candidates in the (top left) low-, (top right) central-, (bottom left) high-$q^2$, and (bottom right) $J/\psi$ bins. 
    The data are summed over the three run periods and are compared with the result of the fit described in the text. 
    }
    \label{supp:fig:muon-fits}
\end{figure}

Figure~\ref{supp:fig:muon-fits} shows the mass distributions of selected \BsPhiMM candidates in the different \qsq bins.
The significant difference in mass resolution between \BsPhiMM and \BsPhiEE decays, where the latter is shown in Fig.~\ref{fig:phill} in the Letter, is due to bremsstrahlung from the \epm and the imperfect nature of the bremsstrahlung recovery.

\clearpage
\section{Differential branching fraction of the \texorpdfstring{\BsPhiEE}{Bs ->phi ee} decay}

\begin{figure}[!htb]
\centering
\includegraphics[width=0.5\linewidth]{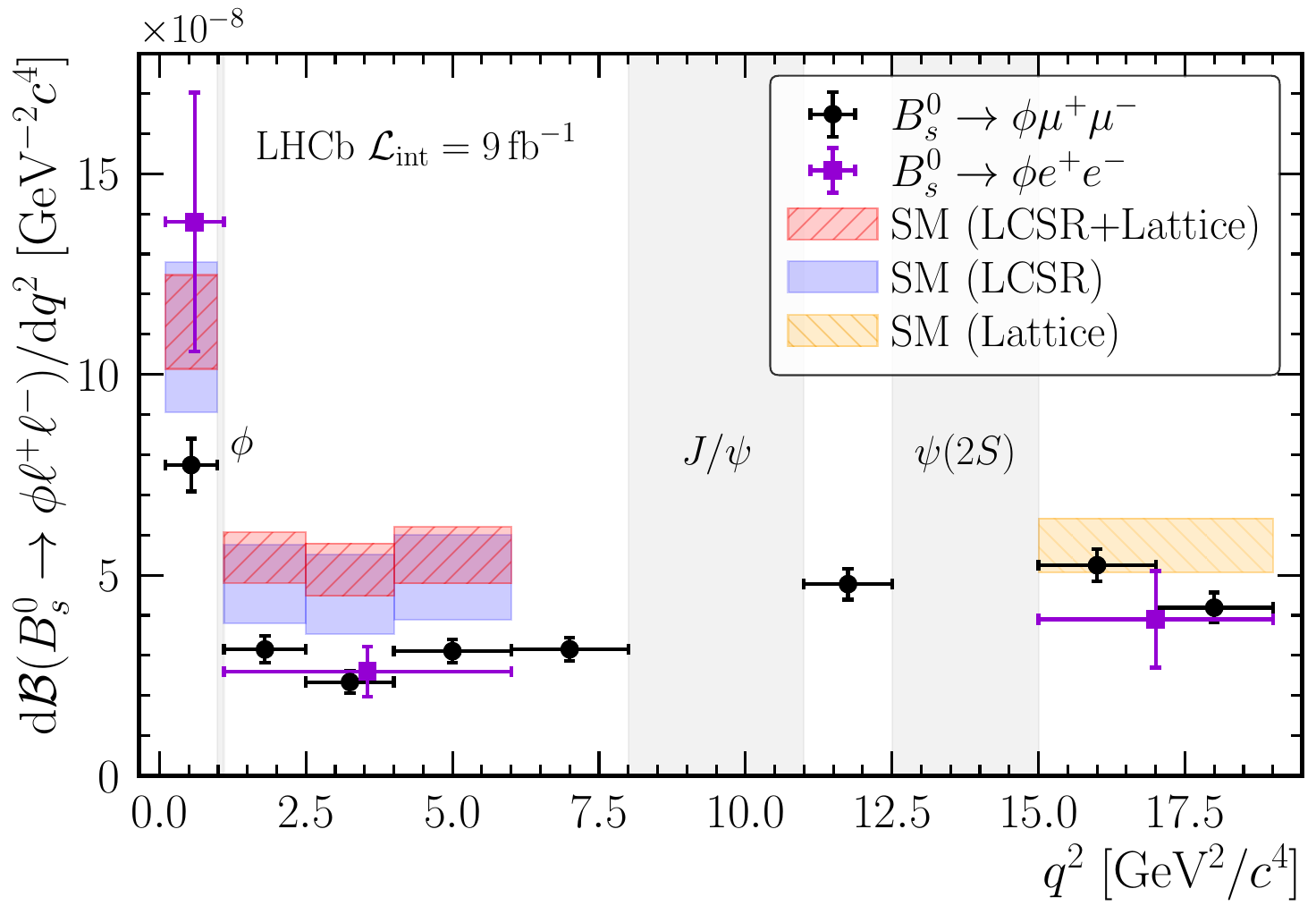}
\caption{ 
Differential branching fraction of the $B_s^0 \rightarrow \phi e^+ e^-$ decay in bins of $q^2$, compared with measurements from $B_s^0 \rightarrow \phi \mu^+\mu^-$ decays in Ref.~\cite{LHCb-PAPER-2021-014} and SM predictions from \textsc{flavio}~\cite{Straub:2018kue} using form factors calculated with light-cone sum-rule~\cite{Bharucha:2015bzk} and lattice QCD techniques~\cite{Horgan:2013pva,Horgan:2015vla}. 
}
\label{fig:summary-branching-fractions}
\end{figure}

Figure~\ref{fig:summary-branching-fractions} presents a summary of the measured \BsPhiEE branching fraction from this letter, compared with that of the \BsPhiMM decay~\cite{LHCb-PAPER-2021-014} and predictions based on the SM extracted from Ref.~\cite{Straub:2018kue} with uncertainties following Refs.~\cite{Bharucha:2015bzk,Horgan:2013pva,Horgan:2015vla}. 

%% file: Authorship_LHCb-PAPER-2024-032.tex
\centerline
{\large\bf LHCb collaboration}
\begin
{flushleft}
\small
R.~Aaij$^{38}$\lhcborcid{0000-0003-0533-1952},
A.S.W.~Abdelmotteleb$^{57}$\lhcborcid{0000-0001-7905-0542},
C.~Abellan~Beteta$^{51}$,
F.~Abudin{\'e}n$^{57}$\lhcborcid{0000-0002-6737-3528},
T.~Ackernley$^{61}$\lhcborcid{0000-0002-5951-3498},
A. A. ~Adefisoye$^{69}$\lhcborcid{0000-0003-2448-1550},
B.~Adeva$^{47}$\lhcborcid{0000-0001-9756-3712},
M.~Adinolfi$^{55}$\lhcborcid{0000-0002-1326-1264},
P.~Adlarson$^{82}$\lhcborcid{0000-0001-6280-3851},
C.~Agapopoulou$^{14}$\lhcborcid{0000-0002-2368-0147}
% ...

{\footnotesize \it

% ...
$^{14}$Universit{\'e} Paris-Saclay, CNRS/IN2P3, IJCLab, Orsay, France\\
% ...
$^{38}$Nikhef National Institute for Subatomic Physics, Amsterdam, Netherlands\\
% ...
$^{47}$Instituto Galego de F{\'\i}sica de Altas Enerx{\'\i}as (IGFAE), Universidade de Santiago de Compostela, Santiago de Compostela, Spain\\
% ...
$^{51}$Physik-Institut, Universit{\"a}t Z{\"u}rich, Z{\"u}rich, Switzerland\\
% ...
$^{55}$H.H. Wills Physics Laboratory, University of Bristol, Bristol, United Kingdom\\
% ...
$^{57}$Department of Physics, University of Warwick, Coventry, United Kingdom\\
% ...
$^{60}$School of Physics and Astronomy, University of Glasgow, Glasgow, United Kingdom\\
$^{61}$Oliver Lodge Laboratory, University of Liverpool, Liverpool, United Kingdom\\
% ...
$^{69}$Syracuse University, Syracuse, NY, United States\\
% ...
$^{82}$Department of Physics and Astronomy, Uppsala University, Uppsala, Sweden, associated to $^{60}$\\
% ...
}
\end{flushleft}